# The Persistence of the Dirty Air Penalty: A Causal Analysis of Formula 1's 2022 Ground Effect Regulations

Bhavay Joshi
Btech Cse
SRMIST
bj4468@srmist.edu.in / bhavayjwork@gmail.com

## Abstract

In early 2022, FIA mandated a redesign of Formula 1 car design focusing on ground effect aerodynamics, with the goal to reduce the performance penalty caused by “Dirty Air” in cases where cars follow each other closely. The claim became part of the regulatory and competitive discourse but was never verified against timing records in relation to individual lap performance, which might reveal the actual performance difference between cars in the official race. We collected over 100486 racing laps and created a dataset for the seasons 2021 to 2025 (2026 race season is in progress) using FastF1 API. We created compound-specific physics correction for Dirty Air to isolate it from fuel and tyre wear confounding effects and evaluated the change in Dirty Air penalty across the 2022 regulation change using two independent techniques - OLS regression with clustered standard errors and Causal Forest Double/Debiased Machine Learning (DML) to account for potential non linearity and effect heterogeneity. We also repeated the analysis of both models with season as a categorical variable to see if any individual season differs significantly from the 2021 baseline in OLS, applying a generalized circuit filter to ensure results weren't skewed by calendar changes.

In both cases, we reached the same conclusion, i.e., that no statistically significant decrease in Dirty Air lap time penalty occurred if 2022 to 2025 were viewed as a pooled period (OLS interaction term $\beta = -0.0025$, $p = 0.803$; Causal Forest mean effects of 0.019 s and 0.018 s per second of following distance in 2021 and 2022 to 2025, respectively). Then we considered a more specific hypothesis motivated by the external aerodynamic simulation data that the decrease was achieved only in 2022 and partially lost in 2023 to 2025. The results of our estimation of both models with season as a categorical variable show that no seasonal effects differ significantly from the baseline of 2021 in OLS, though several seasons can be distinguished from zero in Causal Forest estimates in a non-monotonic pattern and do not match the erosion shape.

We conclude from our findings that there was no significant change in the Dirty Air penalty under the 2022 ground effect rule in the racing data that we analyzed using both the pooled and season level tests. By separating the simulation-based engineering claim from the actual racing outcome data, we demonstrate that even if an aerodynamic advantage is true in theory, it has failed to show any performance advantage in the real racetrack where tyre strategy, driver behavior, and race context can swamp a genuine aerodynamic effect that exists in simulation.

# 1. Introduction

## 1.1 Motivation

In Formula One racing, aerodynamics play a vital role in determining how well a car performs in the race i.e. front and rear wings. And starting from 2022 underbody airflow tunnels provide the downforce that helps the car press against the track and achieve higher cornering speed. Downforce is dependent on the uninterrupted flow of air that flows on to the surface of the car's aerodynamics. If another car follows the leading car very close then it will pass through the disturbed air and experience the low energy wake of this air known as "Dirty Air" and therefore lose a lot of its aerodynamic performance.The real life consequence is well documented in the sport as cars in the dirty air struggle to maintain cornering speed and therefore find it difficult to close the gap for overtaking. This dynamic is a big contributor to the sport's historically low on-track overtaking rates.
In the 2022 season, The FIA introduced a major redesign of their technical regulations. In the new regulations, the principal source of the downforce became ground effect underbody tunnels instead of wing-based aerodynamics. The FIA has claimed that because the downforce produced through ground effects is less affected by the air flows coming from behind another car, this redesign will minimize the negative impact on the performance from close following and therefore will lead to an increase in overtaking attempts. This claim is falsifiable and is easily testable using available race times data.
There are at least two types of evidence relevant to the claim. First are the simulation-based aerodynamic data, and second are the race data (such as real race lap times). This paper is concerned with the latter type of data only. There is no access to any CFD data or team/FIA proprietary aerodynamic data, and therefore there are no claims concerning it. The current study is about testing whether the claimed aerodynamic benefit has produced any measurable result in real races' lap times, a complementary but distinct question from whether the aerodynamic phenomenon actually exists in simulations.

## 1.2 Research Question and Contribution

The following research focuses on 2022 rules on ground effect to identify whether they provided any statistical evidence of the car running behind reducing its lap time penalty for a car running behind another. The subquestion which arises out of this is whether, if there was any evidence of reductions in lap times, they remained consistent between 2022 and 2025 or were different each year.
Our contribution has 4 parts. First, we have created an open-source and reproducible lapwise dataset that spans 5 Formula 1 seasons, and also a methodology for physics correction which isolates Dirty Air effects from the fuel load and tyre degradation. Secondly, we have validated the claim made by FIA with a controlled regression model that helped us address issues such as error correlation and circuit-era collinearity in the data. Thirdly, in order to validate the results obtained from the above model, we have used a Causal Forest Double Machine Learning approach, which helps in detecting hidden variation in the data and complex patterns that normal linear regression baseline models cannot detect. Fourth, we incorporated a seasonal analysis into both of the previous analyses as a test of robustness against an alternative hypothesis that a simple pre/post comparison cannot resolve

### 1.3 Summary of finding

Neither the pooled specification nor the season-level specification found a statistically robust, directionally consistent reduction in the Dirty Air penalty. This holds for the general pre/post comparison and for the year-by-year test based on a specific alternative hypothesis. We compared this result with an external simulation-based study showing a different pattern, and explained why the two types of evidence don't agree.

# 2. Data

### 2.1 Source and Collection

We used FastF1 to collect lap-by-lap data from five seasons of Grand Prix (2021-2025) to get “season”, “circuit”, “driver”, “lap number”, “raw lap time”, “tyre compound”, “TyreLife”, “running position”, “gap to leader”, “GapToCarAhead”, “stint length”. “GapToCarAhead” tells us the time gap between two consecutive cars, whose one whole row represents a lap by the driver in that race.
“GapToCarAhead” is official F1 timing data that tells the time gap between two consecutive cars, and dirty air's aerodynamic wake is conventionally modelled as a function of this time gap, which is usually around 1-2 seconds.
Aerodynamic wake in F1 follows the racing line and does not depend on physical track positioning, and here time gap helps by being the operational measure for exposure in dirty air.

### 2.2 Sample Construction and Composition

We kept only the green flags (TrackStatus == 1) and no pit stops or dry compounds (SOFT, MEDIUM, HARD). All INTERMEDIATE and WET laps were removed as grip and weather changes give lap time variations that are not there from following another car. These laps were irrelevant noise and were dropped. To confirm this was correct, in Section 3.1, we conducted a diagnostic regression which had positive bias in the predicted lap time for intermediate-compound laps, this matched the expectation that wet weather dynamics were producing this contamination.
The final sample size after filtration was 100,486 laps. The sample is imbalanced because 2021 had lesser laps than post 2022, i.e. 20362 laps were from 2021(pre) and 85760 laps were from 2022-2025(post), it is because there were more race disruptions in 2021 like safety cars and season counts, that green flag filter removes disproportionately. Also, circuits like Las Vegas and Miami only exist in later seasons and are not present in 2021 at all.
To counter these issues, circuit fixed effects were introduced and added circuit filtering approaches that are discussed in section 3.4 and extended in 3.6.

# 3. Methodology

## 3.1 Error Modelling and Filtering

We fitted Random Forest Regressions of raw lap time on "TyreLife", "LapNumber", "StintLength", "GapToCarAhead", "GapToLeader" and ran these separately for the 2021 and years after 2022 (post) subsamples. This helped examine the residual structure before building the corrected outcome variable. Baku and São Paulo had substantially higher variance than other circuits. These events had more incidents and so had higher variance. We also found that INTERMEDIATE compound laps have a systematic positive mean bias of approx. 2 seconds, this is what justified the compound restriction in Section 2.2. All of the mean residuals for driver-level effects were under 1 second. Driving style is not a material confounder here.

## 3.2 Lap Time Correction: Isolating the Dirty Air Signal

Raw lap time mixes three unrelated things to aerodynamics, fuel load, tyre degradation, and baseline pace of circuits. To solve this, we made a variable called "CorrectedLapTime", it filters out the first two and leaves only the circuit differences to be filtered in fixed effects at regression stage

**fuel load**: Since fuel load isn't explicitly given, we used lap number in the race as a substitute for the fuel's burning, like most public F1 analytics work

**tyre degradation**: for this, we estimated tyre degradation and fuel load components jointly per compound, via the specification

$$\mathbf{LapTime = \alpha + \beta_1 \cdot TyreLife + \beta_2 \cdot LapNumber + \varepsilon}$$

For tyre degradation, we fit separately per compound (SOFT, MEDIUM, HARD) and then we get the CorrectedLapTime by subtracting TyreLife and LapNumber.
TyreLife was fitted with Lapnumber because at every stint, fresh tyres are installed and at the start of the first stint the fuel is maximum; as the race progresses, the tyre ages making them co-linear with each other.
Also, when we fit them alone, TyreLife changes the degradation coefficient to the wrong sign because it absorbs some part of the fuel effect. To counter this, we added LapNumber as the second regressor to separate out the fuel burn and get a genuine degradation in the TyreLife coefficient

## 3.3 Descriptive Penalty Curve Construction

We used bin gaps of (0, 0.5], (0.5, 1], (1, 2], (2, ∞) seconds on GapToCarAhead. Within each bin, we averaged CorrectedLapTime per circuit, then we averaged those circuit-level averages across circuits by era. This two step method helps to ensure that each circuit is counted equally, regardless of how many laps it has. We can't use it to make inferential or causal claims because it does not control for the differences in circuit composition that happen across eras (i.e., different circuits present in different eras could bias a naive average)

## 3.4 Regression Specification (OLS)

**CorrectedLapTime = $\beta_0$ + $\beta_1$·GapToCarAhead + $\beta_2$·IsPost2022 + $\beta_3$·(GapToCarAhead × IsPost2022) + $\gamma$·Circuit + $\delta$·Compound + $\beta_4$·TyreLife + $\beta_5$·LapNumber + $\varepsilon$**

IsPost2022 is a marker that helps distinguish 2022-2025 seasons and Circuit and Compound help to keep estimates unbiased and serve as categorical fixed effects. The coefficient of primary interest ($\beta_3$) is the interaction term GapToCarAhead × IsPost2022, It measures how the penalty for following another car changes in the new era vs. pre 2022 baseline. The Hypothesis here is: if the regulations worked as intended, we can expect $\beta_3 > 0$, which is a measurable improvement.
Here, we estimated two corrections in the model to make sure the results are actually reliable.
First, since lap times within a single event are serially correlated (one lap affects the next) making the data points dependent, we used standard errors clustered by race (Season × Circuit).
Second, we restricted the primary regression sample to circuits present in both 2021 (pre) and 2022 and after eras (post), because the circuits exclusive to 2022 and later eras (Las Vegas, Miami) make Circuit and IsPost2022 collinear for those circuits.

## 3.5 Causal Forest Double/Debiased Machine Learning

Since OLS assumes the dirty air penalty is constant across all observations, Causal Forest DML helps in the checking for heterogeneous effects. "GapToCarAhead" was the treatment and "CorrectedLapTime" the outcome. Also included "IsPost2022", to check whether the effect differed across eras. After controlling "Circuit", "Compound", "TyreLife", "LapNumber", we modelled Random Forest controls with 100 estimators for each and kept the same circuit restriction as before to keep the comparison equal.

## 3.6 Season Level Robustness Extension

The pooled-tests from Sections 3.4 and 3.5 had 2022-2025 as a single block, that made it catch only the constant effects. This can't differentiate it from a temporary "win" that happened in 2022 and faded by 2025. The rise then fade pattern averaged toward null in the pooled indicator, and was the same as having no effects at all. This is a real concern as the aerodynamic simulator data in Section 5.2 suggests that the benefit was strong in 2022, but later eroded afterwards. For this, we re-estimated both the models with Season as the main variable and captured the actual year-by-year trajectory.
**OLS Change**: Interacted "GapToCarAhead" with C(Season, Treatment(reference=2021)), that gave one coefficient per season (2022, 2023, 2024, 2025) vs. 2021, and loosened the circuit filter to "present in at least two distinct seasons".
**Causal Forest changes**: Interchanged variables from “IsPost2022” to “Season”. Controls and nuisance models stayed unchanged. Also added 95% confidence intervals per season through effect_interval(), that helped close the point-estimate gap

# 4. Results

## 4.1 Corrected Lap Time

The joint TyreLife + LapNumber correction gives these coefficient of the compound:

**Table 1:** Compound specific TyreLife and LapNumber coefficients from the joint tyre degradation/fuel load regression (Section 3.2)

| Compound | TyreLife coefficient (s/lap) | LapNumber coefficient (s/lap) |
|---|---|---|
| HARD | +0.072 | −0.247 |
| MEDIUM | −0.103 | −0.139 |
| SOFT | −0.321 | −0.056 |

HARD tyres are steady and easy to predict as they degrade, so they have a linear degradation. They also start to lose their grip right away and get slower as the lap progresses. In the result, there is a steady and linear degradation of them, this shows up as a positive TyreLife coefficient, with lap time rising consistently as the tyre ages. SOFT and MEDIUM tyres came out moderately negative even when they were estimated after joining. By these results we interpret that the linear degradation is lesser in compounds having non linear degradation (we get into it further in Section 6). This does not affect the causal comparison in Section 4.3, since the same correction is applied to both eras.

## 4.2 Descriptive Penalty Curve

Mean corrected lap time comes out as uniformly higher across all following-distance bins in 2021 (pre) than "after 2022" (post). This tells us that this pattern points to differences in circuit composition between eras and is not a dirty air effect.This is because a real dirty air effect concentrates at close-following bins and does not spread evenly like this. Also both of these eras show a non monotonic (U-shaped) pattern, where the corrected lap time is actually faster in the (0.5, 1] bin than in either the closest bin (0-0.5s) or the cleanest baseline bin (2s+). This is addressed in detail in Section 5.3. GapToCarAhead was independently confirmed to represent a genuine time-based proximity and was not a measurement artifact.

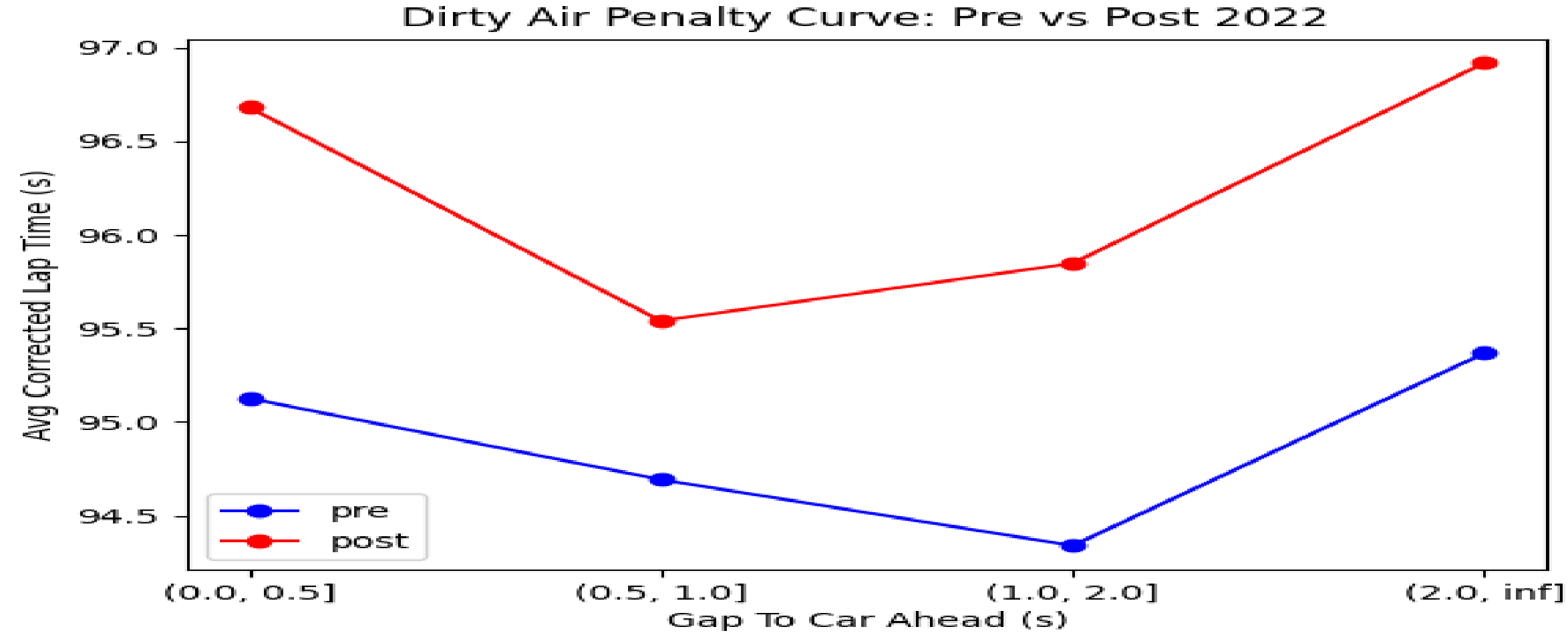

## 4.3 Pooled OLS Comparison Results

**Table 2:** reports coefficients of primary interest, estimated on the sample containing both eras (n = 70,618) with standard errors clustered by race.

| Variable | Coefficient | Cluster robust SE | p-value |
|---|---|---|---|
| GapToCarAhead | 0.0137 | 0.0075 | 0.073 |
| IsPost2022 | −0.293 | 0.279 | 0.296 |
| **GapToCarAhead × IsPost2022** | **−0.0025** | 0.010 | **0.803** |
| TyreLife | 0.082 | 0.011 | < 0.001 |
| LapNumber | 0.087 | 0.007 | < 0.001 |

The Model $R^2 = 0.937$. The interaction coefficient is very small (−0.0025) and statistically the same as zero (0.803). This tells us that there was no evidence that the regulations in 2022 affected the effect of following distance on lap time. The earlier version used the full sample without circuit restriction and had a covariance matrix rank deficiency due to circuit and era collinearity. We restricted the sample to circuits that were present in both eras, shifting the interaction term's p-value from 0.718 to 0.803. So the collinearity was just inflating the confidence in the original unfixed result. The restricted-sample result with the correction is thus more trustworthy and it still doesn't show a statistically detectable effect

## 4.4 Causal Forest Pooled DML Comparison Results

mean estimated treatment effect in 2021 (pre) subsample had 0.0190 seconds per second of following-distance and mean estimated treatment effect of years after 2022 (post) subsample had 0.0181 seconds per second of following-distance. These two values have nearly the same magnitude and direction as the pooled OLS coefficient. The full distribution of lap-level treatment effect in 2021 (pre) and years after 2022 (post) were unimodal. Both distributions closely overlap each other and there was no material difference between the two distributions in terms of shape, spread and location. This shows that the absence of detectable effect is not due to offsetting subgroup effects.

## 4.5 Season Level Robustness Results

**Table 3:** reports the season interaction OLS results (n = 95,720, $R^2 = 0.942$), estimated with the generalized circuit filter described in Section 3.6

| Term | Coefficient | p value |
|---|---|---|
| GapToCarAhead (2021 baseline) | 0.0108 | 0.161 |
| GapToCarAhead × Season[2022] | +0.0153 | 0.344 |
| GapToCarAhead × Season[2023] | −0.0170 | 0.190 |
| GapToCarAhead × Season[2024] | −0.0117 | 0.261 |
| GapToCarAhead × Season[2025] | −0.0140 | 0.157 |

None of the four season-specific interaction terms differ significantly from 2021, nor the baseline itself. This further supports the pooled null in Section 4.3, but at a much finer temporal resolution. When the post-2022 pool was broken down by season, none of the years had a following-distance effect that was convincingly different from 2021.

**Table 4:** reports the season-level Causal Forest mean treatment effects with 95% confidence intervals.

| Season | Mean Effect | 95% confidence interval | Distinguishable from zero |
|---|---|---|---|
| 2021 | 0.0186 | [0.0164, 0.0208] | Yes (Positive) |
| 2022 | 0.0137 | [0.0109, 0.0166] | Yes (Positive) |
| 2023 | 0.0001 | [−0.0046, 0.0048] | No |
| 2024 | 0.0074 | [0.0039, 0.0110] | Yes (Positive) |
| 2025 | −0.0052 | [−0.0088, −0.0016] | Yes (Negative) |

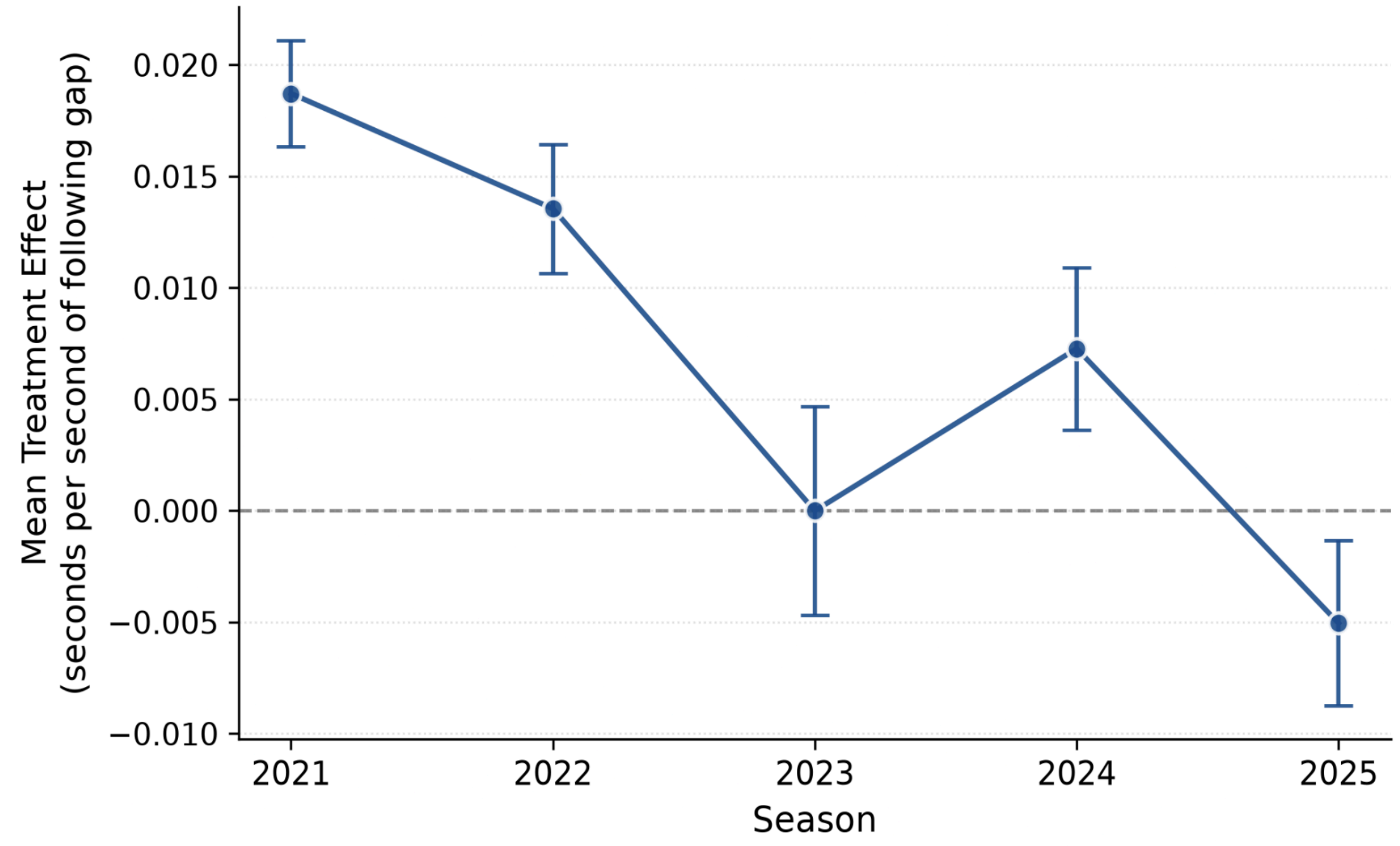

Many individual seasons have non-zero effects. This is a different test than the comparison to the 2021 baseline in Table 3. There are some years that are non-zero in Table 4 but none differ from 2021 in Table 3. Such tension probably indicates limited statistical power at the season level, not a real contradiction. Table 4 uses the same sign convention as the descriptive penalty curve in 4.2: positive effects mean that closer following leads to faster lap times, consistent with the U-shaped pattern in 4.2. The year-on-year pattern is not monotonic and peaks in 2021: positive and large in 2021, smaller but positive in 2022, flattish in 2023, rising to positive again in 2024, and negative in 2025 (the only season on the expected side of zero)

## 4.6 Convergence and Divergence of Results

Both the pooled OLS and Causal Forest analyses agree on the same finding: the dirty air penalty didn't meaningfully change after the 2022 regulations, neither in the pooled pre/post comparison nor at the season level (Section 4.5). The season-level Causal Forest results do add one more nuanced piece: some individual years show non-zero effects, and by 2025 the sign shifts toward the direction you'd expect if the regulations had actually worked. But this shift isn't part of a clean trend, it's irregular, and on its own it doesn't overturn the null finding already established by the season baseline comparison in Table 3.

# 5. Discussion

## 5.1 Interpreting the Null Result

**The key finding from this analysis is that the 2022 ground effect regulations did not lead to any statistically significant sustained reduction in the dirty air penalty, neither in the pooled specification, nor season by season.** This is an important empirical finding as it is a test of a falsifiable public hypothesis against actual outcomes, not merely a confirmation of an expected finding. There are two reasons why this is true. First, controlling for the collinearity issues that exist during the circuit era made the pooled findings even less confident. Second, OLS and Causal Forest DML both agree on the pooled findings and the null is satisfied seasonally as well.

## 5.2 Relationship to Externally Reported Aerodynamic Simulation Data

The downforce loss in the simulation data of the FIA shows a sharp fall in 2022 before going back up over 2023 to 2025 as the teams used different aerodynamics to create a turbulent wake, referred to as "outwash". It is much more specific than just a before/after split and that's precisely why we implemented our seasonal check in Section 3.6 in the first place. The results don't align with this trend. According to Table 4, there isn't any clear decay after reaching its maximum value in 2022, but 2024 breaks the pattern outright. CFD downforce loss percentages and lap times are very different types of measurements. The first one uses a controlled simulation at fixed distances while the latter is affected by tyre strategy, driver behavior and everything else that happens in real races. An aerodynamic effect can be present in simulation but not manifest itself in real-world lap times. This is a gap between two types of evidence, not a refutation of FIA's findings, and we don't have the data to say which one better reflects what's actually happening on track.

## 5.3 The Unresolved U Shape

A pattern that is not fully explained can be seen, but it is present twice, separately. The non-monotonic relationship between following distance and corrected lap time was first noted in section 4.2, and then seen again in the sign pattern of season level Causal Forest in section 4.5. We have an explanation based on driver behavior, where the drivers try harder when they race actively in the middle following distances. However, we have not independently confirmed that it is really what causes the effect, and it seems to be a good research area for the future.

## 6. Limitations

lap number is the proxy for fuel load, not a direct measurement. The green-flag filter removed more 2021 laps, as this year was characterized by more safety cars and interrupted races. Tyre degradation correction still shows a minor bias in case of SOFT and MEDIUM tyres, which have a non-linear wear, not caught by linear adjustment. Unconfoundedness assumption of Causal Forest DML is unverifiable. The seasonal-level modification (3.6) has its disadvantages as well, as only five seasons provide insufficient power and the pattern in Table 4 is likely to be just noise. Comparison in Table 3 (versus 2021) and Table 4 (versus zero) answer different questions and their inconsistency comes from the lack of power, not contradiction. We work with lap time outcomes and not with CFD, and thus comparison in (5.2) is a consistency check rather than a test. Heterogeneity across compounds and drivers occurred diagnostically (3.1), but was not sufficient for the main model and is left for future work

## 7. Conclusion

Using five seasons of F1 timing data and two independent statistical approaches which were tested both pooled and season by season, **we find no evidence that the 2022 ground effect regulations produced a sustained, detectable reduction in the dirty air lap time penalty.** This holds even against a more granular alternative hypothesis based on externally reported aero simulation data showing a different pattern. The distinction between simulation based engineering claims and outcome based race data matters here. Even if a genuine aerodynamic benefit exists in simulation, **it hasn't translated into a detectable gain on the actual racetrack**. Tyre strategy, driver behavior, and race context are enough to swamp a real aero effect once it's tested against real racing conditions rather than controlled CFD output. We're not claiming the underlying phenomenon doesn't exist, we're saying it doesn't show up as a measurable effect in the lap times actually produced on track

## Appendix A: Reproducibility Notes

Data collection, OLS analyses (Python 3.14) (fastf1, pandas, scikit-learn, statsmodels). Causal Forest DML required Python 3.11, since econml doesn't yet support 3.14
https://github.com/BhavayJ-here/The-Persistence-of-the-Dirty-Air-Penalty--Research-Paper